\documentclass[sigconf]{acmart}

\usepackage{booktabs} 

\usepackage{subfigure}
\usepackage{balance}

\copyrightyear{2018}
\acmYear{2018} 
\setcopyright{iw3c2w3}
\acmConference[WWW 2018]{The 2018 Web Conference}{April 23--27, 2018}{Lyon, France}
\acmBooktitle{WWW 2018: The 2018 Web Conference, April 23--27, 2018, Lyon, France}
\acmPrice{}
\acmDOI{10.1145/3178876.3186039}
\acmISBN{978-1-4503-5639-8/18/04.}

\begin{document}
\title{Reinforcement Mechanism Design for e-commerce}

\author{Qingpeng Cai}
\affiliation{%
  \institution{Tsinghua University}
  \city{Beijing} 
  \state{China} }
\email{cqp14@mails.tsinghua.edu.cn}

\author{Aris Filos-Ratsikas}
\affiliation{%
  \institution{University of Oxford}
  \city{Oxford} 
  \state{United Kingdom} 
}
\email{Aris.Filos-Ratsikas@cs.ox.ac.uk}

\author{Pingzhong Tang}
\affiliation{%
	\institution{Tsinghua University}%
		\city{Beijing} 
	\state{China} 
}
\email{kenshinping@gmail.com}

\author{Yiwei Zhang}
\affiliation{%
	\institution{University of California, Berkeley}
	\city{Berkeley} 
	\state{United States of America} 
}
\email{zhangyiwei1234567@126.com}

\renewcommand{\shortauthors}{Cai et al.}

\begin{abstract}
We study the problem of allocating impressions to sellers in e-commerce websites, such as Amazon, eBay or Taobao, aiming to maximize the total revenue generated by the platform. 
We employ a general framework of \emph{reinforcement mechanism design}, which uses deep reinforcement learning to design efficient algorithms, taking the strategic behaviour of the sellers into account. Specifically, we model the impression allocation problem as a Markov decision process, where the states encode the history of impressions, prices, transactions and generated revenue and the actions are the possible impression allocations in each round. To tackle the problem of continuity and high-dimensionality of states and actions, we adopt the ideas of the DDPG algorithm to design an actor-critic policy gradient algorithm which takes advantage of the problem domain in order to achieve convergence and stability. 

We evaluate our proposed algorithm, coined IA(GRU), by comparing it against DDPG, as well as several natural heuristics, under different rationality models for the sellers - we assume that sellers follow well-known no-regret type strategies which may vary in their degree of sophistication. We find that IA(GRU) outperforms all algorithms in terms of the total revenue.
\end{abstract}

%
%



\maketitle

\section{Introduction}

A fundamental problem that all e-commerce websites are faced with is to decide how to allocate the buyer impressions to the potential sellers. When a buyer searches a keyword such as ``iPhone 7 rose gold'', the platform will return a ranking of different sellers providing an item that fits the keyword, with different prices and different historical sale records. The goal of the platform is to come up with algorithms that will allocate the impressions to the most appropriate sellers, eventually generating more revenue from the transactions.

This setting can be modeled as a resource allocation problem over a sequence of rounds, where in each round, buyers arrive, the algorithm inputs the historical records of the sellers and their prices and outputs such an allocation of impressions. The sellers and the buyers carry out their transactions and the historical records are updated.  In reality, most e-commerce websites employ a class of heuristic algorithms, such as collaborative filtering or content based filtering \cite{ricci2011introduction}, many of which rank sellers in terms of ``historical scores'' calculated based on the transaction history of the sellers with buyers of similar characteristics. 

However, this approach does not typically take into account the fact that sellers strategize with the choice of prices, as certain sub-optimal prices in one round might affect the record histories of sellers in subsequent rounds, yielding more revenue for them in the long run. Even worse, since the sellers are usually not aware of the algorithm in use, they might ``explore'' with their pricing schemes, rendering the system uncontrollable at times. It seems natural that a more sophisticated approach that takes all these factors into account should be in place. 

In the presence of strategic or \emph{rational} individuals, the field of \emph{mechanism design} \cite{maskin2008mechanism} has provided a concrete toolbox for managing or preventing the ill effects of selfish behaviour and achieving desirable objectives. Its main principle is the design of systems in such a way that the strategic behaviour of the participants will lead to outcomes that are aligned with the goals of the society, or the objectives of the designer. Cai et al. \cite{cai2016mechanism} tackle the problem of faking transactions and fraudulent seller behaviour in e-commerce using the tools from the field of \emph{mechanism design}.
A common denominator in most of the classical work in economics is that the participants have access to either full information or some distributional estimate of the preferences of others. 
However, in large and constantly evolving systems like e-commerce websites, the participants interact with the environment in various ways, and adjust their own strategies accordingly and dynamically \cite{nekipelov2015econometrics}. In addition to that, their rationality levels are often bounded by either computational or financial constraints, or even cognitive limitations \cite{rubinstein1998modeling}. 

For the reasons mentioned above, a large recent body of work has advocated that other types of agent behaviour, based on \emph{learning} and \emph{exploration}, are perhaps more appropriate for such large-scale online problems encountered in reality \cite{daskalakis2016learning,hartline2015no,lykouris2016learning,hart2000simple,hart2001general,hart2005adaptive,nekipelov2015econometrics,pimachine}. In turn, this generates a requirement for new algorithmic techniques for solving those problems. Our approach is to use techniques from \emph{deep reinforcement learning} for solving the problem of the impression allocation to sellers, given their selfish nature. In other words, given a rationality model for the sellers, we design reinforcement learning algorithms that take this model into account and solve the impression allocation problem efficiently. This general approach is called \emph{reinforcement mechanism design} \cite{tangreinforcement,shen2017reinforcement,cai2018reinforcement}, and we can view our contribution in this paper as an instance of this framework.

\subsection*{No-regret learning as agent rationality}

As mentioned earlier, the strong informational assumptions of classical mechanism design are arguably unrealistic in complex and dynamic systems, like diverse online marketplaces. Such repeated game formulations typically require that the participants know the \emph{values} of their competitors (or that they can estimate them pretty accurately based on known prior distributions) and that they can compute their payoff-maximizing strategies over a long sequence of rounds. Such tasks are usually computationally burdensome and require strong cognitive assumptions, as the participants would have to reason about the future, against all possible choices of their opponents, and in a constantly evolving environment.

Given this motivation, an alternative approach in the forefront of much of the recent literature in algorithmic mechanism design is to assume that the agents follow some type of \emph{no-regret strategies}; the agent picks a probability mixture over actions at each round and based on the generated payoffs, it updates the probabilities accordingly, minimizing the long-term regret. This is more easily conceivable, since the agents only have to reason about their own strategies and their interaction with the environment, and there is a plethora of no-regret algorithms at their disposal. Precisely the same argument has been made in several recent works \cite{nekipelov2015econometrics,daskalakis2016learning,hartline2015no,lykouris2016learning,pimachine} that study popular auction settings under the same rationality assumptions of no-regret, or similar types. In fact, there exist data from Microsoft's Ad Actions which suggest that advertisers do use no-regret algorithms for their actions \cite{tardostalk}. For a more detailed discussion on related rationality models, the reader is referred to \cite{hart2005adaptive}.\\

\noindent \textbf{The seller rationality model:} To model the different sophistication levels of sellers, we consider four different models of rationality, based on well-established no-regret learning approaches. The first two, \emph{$\epsilon$-Greedy} \cite{watkins1989learning} and \emph{$\epsilon$-First} are known as \emph{semi-uniform methods}, because they maintain a distinction between exploration and exploitation. The later is often referred to as ``A/B testing'' and is widely used in practice \cite{burtini2015survey,chawla2016b}. The other two approaches, \emph{UCB1} \cite{agrawal1995sample,auer1995gambling} and \emph{Exp3} \cite{auer1995gambling,auer2002nonstochastic} are more sophisticated algorithms that differ in their assumptions about the nature of the rewards, i.e. whether they follow unknown distributions or whether they are completely adversarial. Note that all of our rationality models employ algorithms for the \emph{multi-arm bandit} setting, as in platforms like Taobao or eBay, the impression allocation algorithms are unknown to the sellers and therefore they can not calculate the payoffs of unused auctions. The update of the weights to the strategies is based solely on the observed payoffs, which is often referred to as the \emph{bandit feedback} setting \cite{foster2016learning}.

We note here that while other related rationality models can be used, the goal is to choose a model that \emph{real sellers would conceivably use in practice}. The semi-uniform algorithms are quite simpler and model a lower degree of seller sophistication, whereas the other two choices correspond to sellers that perhaps put more effort and resources into optimizing their strategies - some examples of sophisticated optimization services that are being used by online agents are provided in \cite{lykouris2016learning}. Note that both UCB1 and Exp3 are very well-known \cite{burtini2015survey} and the latter is perhaps the most popular bandit feedback implementation of the famous \emph{Hedge} (or \emph{Multiplicative Weights Update}) algorithm for no-regret learning in the fully informed feedback setting.

\subsection*{The impression allocation problem}

We model the impression allocation problem as a Markov decision process (MDP) in which the information about the prices, past transactions, past allocations of impressions and generated revenue is stored in the states, and the actions correspond to all the different ways of allocating the impressions, with the rewards being the immediate revenue generated by each allocation. Given that the costs of the sellers (which depend on their production costs) are private information, it seems natural to employ reinforcement learning techniques for solving the MDP and obtain more sophisticated impression allocation algorithms than the heuristics that platforms currently employ.

In our setting however, since we are allocating a very large number of impressions, both the state space and the action space are extremely large and high-dimensional, which renders traditional reinforcement learning techniques such as temporal difference learning \cite{sutton1988learning} or more specifically Q-learning \cite{dayan1992q} not suitable for solving the MDP. In a highly influential paper, Mnih et al. \cite{mnih2015human} employed the use of deep neural networks as function approximators to estimate the action-value function. The resulting algorithm, coined ``Deep Q Network'' (DQN), can handle large (or even continuous) state spaces but crucially, it can not be used for large or continuous action domains, as it relies on finding the action that maximizes the Q-function at each step.

To handle the large action space, policy gradient methods have been proposed in the literature of reinforcement learning with actor-critic algorithms rising as prominent examples \cite{sutton1999policy,bhatnagar2007incremental, degris2012model}, where the critic estimates the Q-function by exploring, while the actor adjusts the parameters of the policy by stochastic gradient ascent. To handle the high-dimensionality of the action space, Silver et al. \cite{silver2014deterministic} designed a deterministic actor-critic algorithm, coined ``Deterministic Policy Gradient'' (DPG) which performs well in standard reinforcement-learning benchmarks such as mountain car, pendulum and 2D puddle world. As Lillicrap et al. \cite{lillicrap2015continuous} point out however, the algorithm falls short in large-scale problems and for that reason, they developed the ``Deep-DPG'' (DDPG) algorithm which uses the main idea from \cite{mnih2015human} and combines the deterministic policy gradient approach of DPG with deep neural networks as function approximators. To improve convergence and stability, they employ previously known techniques such as batch normalization \cite{ioffe2015batch}, target Q-networks \cite{mnih2013playing}, and experience replay \cite{adam2012experience,heess2015memory,mnih2015human}. \\

\noindent \textbf{The IA(GRU) algorithm:} We draw inspiration from the DDPG algorithm to design a new actor-critic policy gradient algorithm for the impression allocation problem, which we refer to as the $\emph{IA(GRU)}$ algorithm. IA(GRU) takes advantage of the domain properties of the impression allocation problem to counteract the shortcomings of DDPG, which basically lie in its convergence when the number of sellers increases. The modifications of IA(GRU) to the actor and critic networks reduce the policy space to improve convergence and render the algorithm robust to settings with variable sellers, which may arrive and depart in each round, for which DDPG performs poorly. We evaluate IA(GRU) against DDPG as well as several natural heuristics similar to those usually employed by the online platforms and perform comparisons in terms of the total revenue generated. We show that IA(GRU) outperforms all the other algorithms for all four rationality models, as well as a combined pool of sellers of different degrees of sophistication.

\section{The setting}\label{sec:prelims}
In the impression allocation problem of e-commerce websites, there are $m$ sellers who compete for a unit of buyer impression.\footnote{Since the buyer impressions to be allocated is a huge number, we model it as a continuous unit to be fractionally allocated. Even if we used a large integer number instead, the traditional approaches like DDPG fall short for the same reasons and furthermore all of the performance guarantees of IA(GRU) extend to that case.} In each round, a buyer\footnote{As the purchasing behavior is determined by the valuation of buyers over the item, without loss of generality we could consider only one buyer at each round.} searches for a keyword and the platform returns a ranking of sellers who provide an item that matches the keyword; for simplicity, we will assume that all sellers provide identical items that match the keyword exactly. Each seller $i$ has a private cost $c_i$ for the item, which can be interpreted as a production or a purchasing cost drawn from an i.i.d. distribution $F_s$. 

Typically, there are $n$ slots (e.g. positions on a webpage) to be allocated and we let $x_{ij}$ denote the probability (or the fraction of time) that seller $i$ is allocated the impression at slot $j$. With each slot, there is an associated click-through-rate $\alpha_j$ which captures the ``clicking potential'' of each slot, and is independent of the seller, as all items offered are identical. We let $q_i=\sum_{j=1}^{n}x_{ij}\alpha_{j}$ denote the probability that the buyer will click the item of seller $i$. Given this definition (and assuming that sellers can appear in multiple slots in each page), the usual feasibility constraints for allocations, i.e. for all $i$, for all $j$, t holds that $0\leq x_{ij}\leq 1$ and or all $j$, it holds that $\sum_{i=1}^{m}x_{ij}\leq 1$ can be alternatively written as \[\text{for all } i, q_i\geq 0, \text{ it holds that } \sum_{i=1}^{m}q_i\leq \sum_{j=1}^{n}\alpha_j \text{ and }  \sum_{j=1}^{n}\alpha_j=1.\] That is, for any such allocation $q$, there is a feasible ranking $x$ that realizes $q$ (for ease of notation, we assume that the sum of click-through rates of all slots is $1$) and therefore we can allocate the buyer impression to sellers directly instead of outputting a ranking over these items when a buyer searches a keyword.\footnote{The framework extends to cases where we need return similar but different items to a buyer, i.e, the algorithm outputs a ranking over these items. Furthermore, our approach extends trivially to the case when sellers have multiple items.} 

Let $h_{it}=(v_{it},p_{it},n_{it},\ell_{it})$ denote the \emph{records} of seller $i$ at round $t$, which is a tuple consisting of the following quantities: 
\begin{enumerate}
	\item $v_{it}$ is the expected fraction of impressions that seller $i$ gets, 
	\item $p_{it}$ is the price that seller $i$ sets, 
	\item $n_{it}$ is the expected amount of transactions that seller $i$ makes,
	\item $\ell_{it}$ is the expected revenue that seller $i$ makes at round $t$. 
\end{enumerate}
Let $H_t=(h_{1t},h_{2t},...,h_{it})$ denote the records of all sellers at round $t$, and let $H_{it}=(h_{i1},h_{i2},...,h_{it})$ denote the vectors of records of seller $i$ from round $1$ to round $t$, which we will refer to as the \emph{records} of the seller. At each round $t+1$, seller $i$ chooses a price $p_{i(t+1)}$ for its item and the algorithm allocates the buyer impression to sellers. \\

\noindent \textbf{MDP formulation:} The setting can be defined as a \emph{Markov decision process} (MDP) defined by the following components: a continuous state space $\mathcal{S}$, a continuous action space $\mathcal{A}$, with an initial state distribution with density $p_0(s_0)$, and a transition distribution of states with conditional density $p(s_{t+1}|s_{t},a_{t})$ satisfying the Markov property, i.e.  $p(s_{t+1}|s_0,a_0,...,s_t,a_t)=p(s_{t+1}|s_{t},a_{t})$. Furthermore, there is an associated reward function $r:\mathcal{S} \times \mathcal{A}\rightarrow \mathcal{R}$ assigning payoffs to pairs of states and actions. Generally, a policy is a function $\pi$ that selects stochastic actions given a state, i.e, $\pi:\mathcal{S}\rightarrow \mathcal{P}(\mathcal{A})$, where $\mathcal{P}(\mathcal{A})$ is the set of probability distributions on $\mathcal{A}$. Let $R_t$ denote the discounted sum of rewards from the state $s_t$, i.e, $R_t(s_t)=\sum_{k=t}^{\infty}{\gamma}^{k-t}r(s_k,a_k)$, where $0<\gamma<1$. Given a policy and a state, the \emph{value function} is defined to be the expected total discounted reward, i.e. $V^{\pi}(s)=E[R_t(s_t)|s_t=s;\pi]$ and the \emph{action-value function} is defined as $Q^{\pi}(s,a)=E[R_t(s_t)|s_t=s,a_t=a;\pi]$.

For our problem, a state $s_t$ of the MDP consists of the records of all sellers in the last $T$ rounds, i.e. $s_t=(H_{t-T},...,H_{t-1})$, that is, the state is a $(T,m,4)$ tensor, the allocation outcome of the round is the action, and the immediate reward is the expected total revenue generated in this round. The performance of an algorithm is defined as the average expected total revenue over a sequence of $T_0$ rounds. \\

\noindent \textbf{Buyer Behaviour:} We model the behaviour of the buyer as being dependent on a valuation that comes from a distribution with cumulative distribution function $F_b$. 
Intuitively, this captures the fact that buyers may have different spending capabilities (captured by the distribution). 
Specifically, the probability that the buyer purchases item $i$ is $n_{it}=(1-F_b(p_{it}))\cdot v_{it}$, that is, the probability of purchasing is decided by the impression allocation and  the price seller $i$ sets. 
For simplicity and without loss of generality with respect to our framework, we assume that the buyer's valuation is drawn from $U(0,1)$, i.e. the uniform distribution over $[0,1]$.

\subsection*{Seller Rationality}

As we mentioned in the introduction, following a large body of recent literature, we will assume that the sellers employ no-regret type strategies for choosing their prices in the next round. Generally, a seller starts with a probability distribution over all the possible prices, and after each round, it observes the payoffs that these strategies generate and adjusts the probabilities accordingly. As we already explained earlier, it is most natural to assume strategies in the \emph{bandit feedback} setting, where the seller does not observe the payoffs of strategies in the support of its strategy which were not actually used. The reason is that even if we assume that a seller can see the prices chosen in a round by its competitors, it typically does not have sufficient information about the allocation algorithm used by the platform to calculate the payoffs that other prices would have yielded. Therefore it is much more natural to assume that the seller updates its strategy based on the \emph{observed rewards}, using a multi-arm bandit algorithm.

More concretely, the payoff of a seller $i$ that receives $v_{it}$ impressions in round $t$ when using price $p_{ij}(t)$, is given by $u_{ij}(t)=n_{it}(p_{ij}(t)-c_i)=v_{it}(1-F_b(p_{it}))(p_{ij}(t)-c_i)$. For consistency, we normalize the costs and the prices to lie in the unit interval $[0,1]$ and we discretize the price space to a ``dense enough'' grid (of size $1/K$, for some large enough $K$). This discretization can either be enforced by the platform (e.g. the sellers are required to submit bids which are multiples of $0.05$) or can be carried out by the sellers themselves in order to be able to employ the multi-arm bandit algorithms which require the set of actions to be finite, and since small differences in prices are unlikely to make much difference in their payoffs. \\

\noindent We consider the following possible strategies for the sellers, based on well-known bandit algorithms.\\

\noindent \emph{\textbf{$\varepsilon$-Greedy \cite{watkins1989learning}:}} With probability $\varepsilon$, each seller selects a strategy uniformly at random and with
probability $1-\varepsilon$, the strategy with the best observed (empirical) mean payoff so far. The parameter $\varepsilon$ denotes the degree of \emph{exploration} of the seller, whereas $1-\varepsilon$ is the degree of \emph{exploitation}; here $\varepsilon$ is drawn i.i.d. from the normal distribution $N(0.1,0.1/3)$.\\

\noindent \emph{\textbf{$\varepsilon$-First:}} For a horizon of $\mathcal{T}$ rounds, this strategy consists of an \emph{exploration} phase first, over $\varepsilon \cdot \mathcal{T}$ rounds, followed by an \emph{exploitation} phase, for the remaining period. In the exploration phase, the seller picks a strategy uniformly at random. In the remaining rounds, the sellers picks the strategy that maximizes the empirical mean of the observed rewards. For each seller, we set $\mathcal{T}=200$ and $\varepsilon=0.1$.\\

\noindent \emph{\textbf{Exponential-weight Algorithm for Exploration and Exploitation (Exp3) \cite{auer2002nonstochastic,auer1995gambling}:}} We use the definition of the algorithm from \cite{auer2002nonstochastic}. Let $\gamma \in (0,1]$ be a real number and initialize $w_i(1)=1$ for $i=1,\ldots,K+1$ to be the initial weights of the possible prices\footnote{For ease of notation, we drop the subscript referring to a specific seller, as there is no ambiguity.}. In each round $t$,
\begin{itemize}
	\item For $i=1,\ldots,K+1$, let $\pi_i(t)= (1-\gamma) \frac{w_i(t)}{\sum_{j=1}^{K+1} w_j(t)} + \frac{\gamma}{K+1}$, where $w_i(t)$ is the weight of price $p_i$ in round $t$.
	\item Select a price $p_{j}(t)$ according to the probability distribution defined by $\pi_1(t),\ldots,\pi_{K+1}(t)$.
	\item Receive payoff $u_{j}(t) \in [0,1]$.
	\item For $\ell=1,\ldots,K+1$, let
	\[
	\hat{u}_{\ell}(t)=
	\begin{cases}
	u_{\ell}(t)/\pi_{\ell}(t), & \text{if } \ell=j \\
	0, & \text{otherwise} \\
	\end{cases}
	\]
	and $w_{\ell}(t+1) = w_{\ell}(t)e^{\gamma \cdot \hat{u}_{\ell}(t)/(K+1)}$.
\end{itemize}
We remark here that since the payoff of each seller in some round $t$ actually takes values in $[-1,1]$, we scale the payoff to $[0,1]$ by applying the transformation $f(u)=(u+1)/2$ to any payoff $u$.\\

\noindent \emph{\textbf{Upper Confidence Bound Algorithm (UCB1) \cite{agrawal1995sample,auer2002finite}:}}  For each price $p_j \in [0,1/K,2/K,\ldots,1]$, initialize $x_j(1)=0$. At the end of each round $t$, update $x_j(t)$ as:
	\[
x_j(t)=
\begin{cases}
x_j(t-1)/t + u_j(t)/t, & \text{if } j \text{ was chosen in this round } t \\
x_j(t-1), & \text{otherwise} \\
\end{cases}
\]
For any round $t \in \{0,\ldots,K\}$, the seller chooses a price $p_j$ that has not been used before in any of the previous rounds (breaking ties arbitrarily). For any round $t \geq K+1$, the seller chooses the price $p_j$ with the maximum weighted value $x_j$, i.e, 
\[
p_j(t) \in \arg\max_{j \in \{0,1/K,\ldots,1\}} x_j(t)+\frac{\log_2 t}{\sum_{\tau=1}^{t}I_{j\tau}}
\], where $I_{j\tau}$ is the indicator function, i.e.
\[
I_{j\tau}=
\begin{cases}
	1, & \text{if } p_j \text{ was chosen in round } \tau \\
	0, & \text{otherwise.} \\
\end{cases}
\] 
$\varepsilon$-Greedy and $\varepsilon$-First are simple strategies that maintain a clear distinction between exploration and exploitation and belong to the class of semi-uniform strategies. Exp3 is the most widely used bandit version of perhaps the most popular no-regret algorithm for the full information setting, the Hedge (or Multiplicative Weight updates) algorithm \cite{freund1995desicion} and works in the \emph{adversarial} bandit feedback model \cite{auer2002nonstochastic}, where no distributional assumptions are being made about the nature of the rewards. UCB1, as the name suggests, maintains a certain level of optimism towards less frequently played actions (given by the second part of the sum) and together with this, it uses the empirical mean of observed actions so far to choose the action in the next round. The algorithm is best suited in scenarios where the rewards do follow some distribution which is however unknown to the seller. 

For a more detailed exposition of all these different algorithms, \cite{burtini2015survey} provides a concise survey. The point made here is that these choices are quite sensible as they (i) constitute choices that a relatively sophisticated seller, perhaps with a research team at its disposal could make, (ii) can model sellers with different degrees of sophistication or pricing philosophies and (iii) are consistent with the recent literature on algorithmic mechanism design, in terms of modeling agent rationality in complex dynamic environments.

\section{Allocation algorithms}\label{sec:algorithms}

In this section, we will briefly describe the algorithms that we will be comparing IA(GRU) against - two natural heuristics similar to those employed by platforms for the impression allocation problem, as well as the DDPG algorithm of Lillicrap et al. \cite{lillicrap2015continuous}. 

\subsection*{Heuristic Allocation Algorithms}
As the strategies of the sellers are unknown to the platform, and the only information available is the sellers' historical records, the platform can only use that information for the allocation. Note that these heuristics do not take the rationality of the sellers into account, when deciding on the allocation of impressions.\\

\noindent The first algorithm is a simple greedy algorithm, which allocates the impressions proportionally to the revenue contribution.\\

\noindent\textbf{\emph{Greedy Myopic Algorithm}:} At round $0$, the algorithm allocates a $1/m$-fraction of the buyer impression to each seller. At any other round $\tau+1$ (for $\tau\geq 0$),  the algorithm allocates a fraction of $\ell_{i\tau}/\sum_{j=1}^{m}{\ell_{j\tau}}$ of the buyer impression to each seller, i.e. proportionally to the contribution of each seller to the total revenue of the last round.\\

\noindent The second algorithm is an algorithm for the \emph{contextual multi-arm bandit} problem, proposed by \cite{li2010contextual}, based on the principles of the family of upper confidence bound algorithms (UCB1 is an algorithm in this family). The algorithm is among the state of the art solutions for recommender systems \cite{burtini2015survey} and is an example of contextual bandit approaches, which are widely applied to such settings \cite{agrawal2013thompson,krause2011contextual,bouneffouf2012contextual,li2010contextual}.
To prevent any confusion, we clarify here that while we also used bandit algorithms for the seller rationality models, the approach here is fundamentally different as the Linear UCB Algorithm is used for the allocation of impressions - not the choice of prices - and the arms in this case are the different sellers.\\

\noindent\textbf{\emph{Linear UCB Algorithm \cite{li2010contextual}}:} We implement the algorithm as described in \cite{li2010contextual} - in the interest of space, we do not provide the definition of the algorithm, but refer the reader to \emph{Algorithm 1} in \cite{li2010contextual}. We model each seller as an \emph{arm} and set $h_{it}$ as the feature of each arm $i$ in each round $t$. The parameter $\alpha$ is set to $1$.

\subsection*{Deep Deterministic Policy Gradient}
Here, we briefly describe the DDPG algorithm of \cite{lillicrap2015continuous}, which we we draw inspiration from in order to design our impression allocation algorithm. Before describing the algorithm, we briefly mention the main ingredients of its predecessor, the DPG algorithm of Silver et al. \cite{silver2014deterministic}.\\
 
 \noindent\textbf{\emph{Deterministic Policy Gradient}:} The shortcoming of DQN \cite{mnih2015human} is that while it can handle continuous states, it can not handle continuous actions or high-dimensional action spaces. Although stochastic actor-critic algorithms could handle continuous actions, they are hard to converge in high dimensional action spaces. The DPG algorithm \cite{silver2014deterministic} aims to train a deterministic policy $\mu_{\theta}:\mathcal{S}\rightarrow \mathcal{A}$ with parameter vector $\theta \in R^{n}$. This algorithm consists of two components: an actor, which adjusts the parameters $\theta$ of the deterministic policy $\mu_{\theta}(s)$ by stochastic gradient ascent of the gradient of the discounted sum of rewards, and the critic, which approximates the action-value function.\\
 
 \noindent \textbf{\emph{Deep Deterministic Policy Gradient}:} Directly training neural networks for the actor and the critic of the DPG algorithm fails to achieve convergence; the main reason is the high degree of temporal correlation which introduces high variance in the approximation of the Q-function by the critic. For this reason, the DDPG algorithm uses a technique known as \emph{experience replay}, according to which the experiences of the agent at each time step are stored in a replay buffer and then a mini-batch is sampled uniformly at random from this set for learning, to eliminate the temporal correlation. The other modification is the employment of \emph{target networks} for the regularization of the learning algorithm. The target network is used to update the values of $\mu$ and $Q$ at a slower rate instead of updating by the gradient network; the prediction $y_t$ will be relatively fixed and violent jitter at the beginning of training is absorbed by the target network. A similar idea appears in \cite{van2016deep} with the form of double Q-value learning.

\section{The Impression Allocation (GRU) algorithm}\label{sec:algo}

In this section, we present our main deep reinforcement learning algorithm, termed IA(GRU) (``IA'' stands for ``impression allocation'' and ``GRU'' stands for ``gated recurrent unit'') which is in the center of our framework for impression allocations in e-commerce platforms and is based on the ideas of the DDPG algorithm. Before we present the algorithm, we highlight why simply applying DDPG to our problem can not work.\\

\noindent \textbf{Shortcomings of DDPG}: First of all, while DDPG is designed for settings with continuous and often high-dimensional action spaces, the blow-up in the number of actions in our problem is very sharp as the number of sellers increases; this is because the action space is the set of all feasible allocations, which increases very rapidly with the number of sellers. As we will show in Section \ref{sec:experiments}, the direct application of the algorithm fails to converge even for a moderately small number of sellers. The second problem comes from the inability of DDPG to handle variability on the set of sellers. Since the algorithm uses a two-layer fully connected network, the position of each seller plays a fundamental role; each seller is treated as a different entity according to that position. As we show in Section \ref{sec:experiments}, if the costs of sellers at each round are randomly selected, the performance of the DDPG algorithm deteriorates rapidly. The settings in real-life e-commerce platforms however are quite dynamic, with sellers arriving and leaving or their costs varying over time, and for an allocation algorithm to be applicable, it should be able to handle such variability. We expect that each seller's features are only affected by its historical records, not some ``identity'' designated by the allocation algorithm; we refer to this highly desirable property as "permutation invariance". Based on time-serial techniques, our algorithm uses Recurrent Neural Networks at the dimension of the sellers and achieves the property.\\


\noindent \textbf{\emph{The IA(GRU) algorithm}:} 
Next, we explain the design of our algorithm, but we postpone some implementation details for Section \ref{sec:experiments}. At a high level, the algorithm uses the framework of DDPG with different network structures and different inputs of networks. It maintains a sub-actor network and a sub-critic network for each seller and employs \emph{input preprocessing} at each training step, to ensure permutation invariance.\\

\noindent \textbf{Input Preprocessing:} In each step of training, with a state tensor of shape $(T,m,4)$, we firstly utilize a \emph{background network} to calculate a public vector containing information of all sellers: it transforms the state tensor to a $(T,m\times 4)$ tensor and performs RNN operations on the axis of rounds.  At this step, it applies a \emph{permutation transformation}, i.e. a technique for maintaining permutation invariance. Specifically, it first orders the sellers according to a certain metric, such as the weighted average of their past generated revenue and then inputs the (state, action) pair following this order to obtain the public vector $(pv)$.  On the other hand, for each seller $i$, it applies a similar RNN operation on its history, resulting in an individual temporal feature called $(f_i)$. Combining those two features, we obtain a feature vector $(pv,f_i)$ that we will use as input for the sellers' sub-actor and sub-critic networks.\\

\begin{figure}
	\centering
	\includegraphics[scale=0.35]{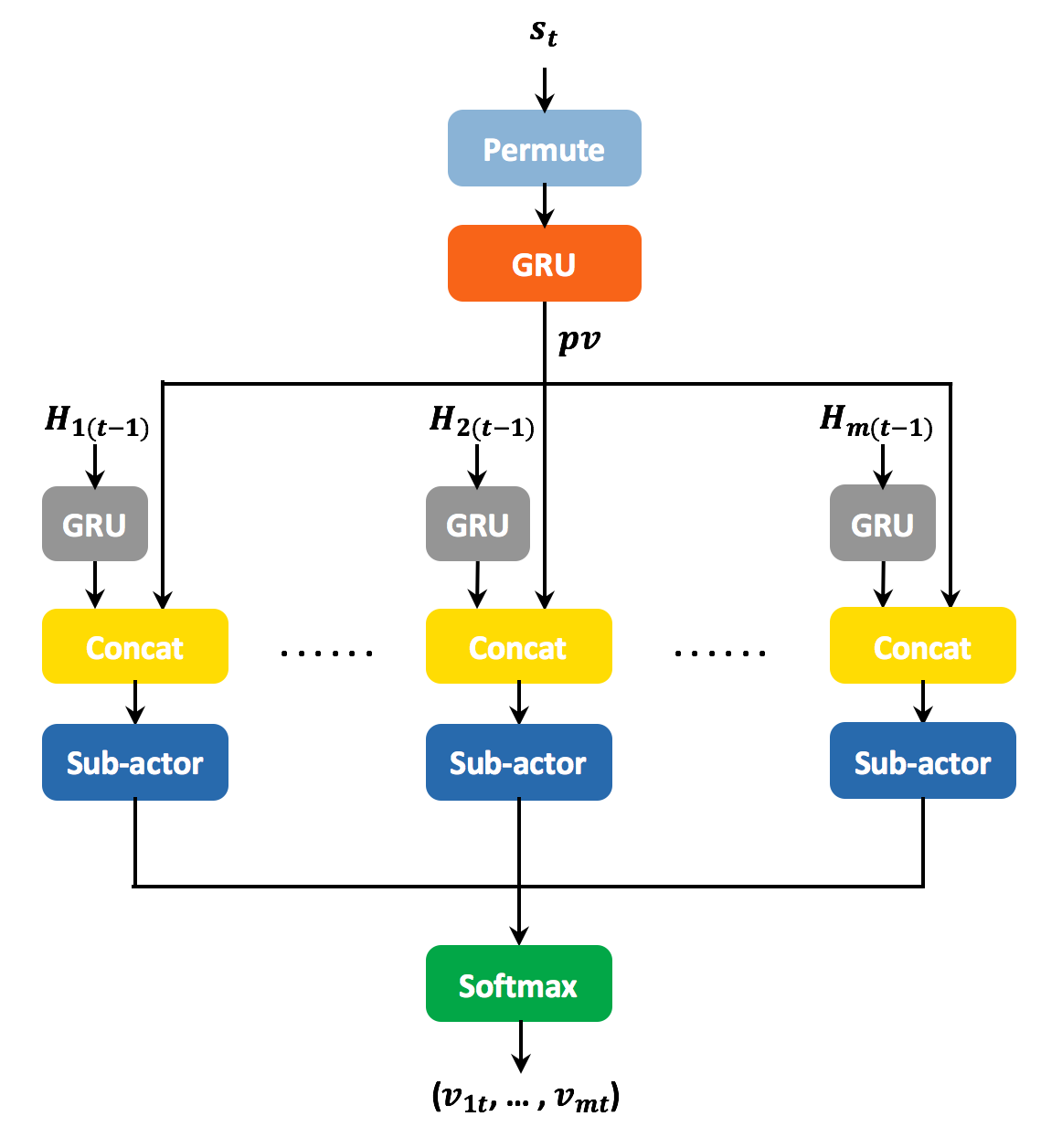}
	\caption{The framework of the actor network of the IA(GRU) algorithm.}
	\label{fig:1}
\end{figure}

\noindent \textbf{Actor network:} For each seller, the input to the sub-actor network is $(pv,f_i)$ and the output is a score. This algorithm uses a softmax function over the outputs of all sub-actor networks in order to choose an action. The structure of the policy which is shown in Figure \ref{fig:1} ensures that the policy space is much smaller than that of DDPG as the space of inputs of all sub-actor networks is restricted, and allows for easier convergence, as we will show in Section \ref{sec:experiments}.\\

\noindent \textbf{Critic network:} For the critic, we make use of a domain-specific property, namely that the immediate reward of each round is the sum of revenues of all sellers and the record of each seller has the same space. Each sub-critic network inputs the expected fraction of buyer impression the seller gets (the sub-action) and $(pv,f_i)$ (the sub-state) as input and outputs the Q-value of the corresponding seller, i.e, the expected discounted sum of revenues from the sub-state following the policy. Then, it sums up the estimated Q-value of all sub-critic networks to output the final estimated Q-value, with the assumption that the strategy of each seller is independent of the records of other sellers, which is the case in all of our rationality models. The framework of the critic network is similar to Figure \ref{fig:1}.

\section{Experimental Evaluation}\label{sec:experiments}

In this section, we present the evaluation of our algorithms in terms of convergence time and revenue performance against several benchmarks, namely the direct application of the DDPG algorithm (with a fully connected network) and the heuristic allocation algorithms that we defined in Section \ref{sec:algorithms}. We use Tensorflow and Keras as the engine for the deep learning, combining the idea of DDPG and the techniques mentioned in Section \ref{sec:algo}, to train the neural network.\\

\noindent \textbf{Designed experiments:} First, we will compare IA(GRU) and DDPG in terms of their convergence properties in the training phase and show that the former converges while the latter does not. Next, we will compare the four different algorithms (Greedy Myopic, Linear UCB, DDPG and IA(GRU)) in terms of the generated revenue for two different settings, a setting with \emph{fixed sellers} and a setting with \emph{variable sellers}. The difference is that in the former case, we sample the costs $c_i$ once in the beginning whereas in the latter case, the cost $c_i$ of each seller is sampled again in each round. This can either model the fact that the production costs of sellers may vary based on unforeseeable factors or simply that sellers of different capabilities may enter the market in each round.

For each one of these two settings, we will compare the four algorithms for each one of the four different rationality models ($\epsilon$-Greedy, $\epsilon$-First, UCB1 and Exp3) \emph{separately} as well as in a \emph{combined} manner, by assuming a mixed pool of sellers, each of which may adopt a different rationality model from the ones above. The latter comparison is meant to capture cases where the population of sellers is heterogeneous and may consist of more capable sellers that employ their R\&D resources to come up with more sophisticated approaches (such as UCB1 or Exp3) but also on more basic sellers that employ simpler strategies (such as $\epsilon$-Greedy). Another interpretation is that the distinction is not necessarily in terms of sophistication, but could also be due to different market research, goals, or general business strategies, which may lead to different decisions in terms of which strategy to adopt.

Our experiments are run for $200$ sellers, a case which already captures a lot of scenarios of interest in real e-commerce platforms. A straightforward application of the reinforcement learning algorithms for much larger numbers of sellers is problematic however, as the action space of the MDP increases significantly, which has drastic effects on their running time. To ensure scalability, we employ a very natural heuristic, where we divide the impression allocation problem into sub-problems and then solve each one of those in parallel. We show at the end of the section that this ``scale-and-solve'' version of IA(GRU) clearly outperforms the other algorithms for large instances consisting of as many as 10.000 sellers.
\\

\noindent \textbf{Experimental Setup:} In the implementation of DDPG, the actor network uses two full connected layers, a rectified linear unit (ReLu) as the activation function, and outputs the action by a softmax function. The critic network inputs a (state,action) pair and outputs the estimation of the Q-value using similar structure. The algorithm IA(GRU) uses the same structure, i.e. the fully connected network in the sub-actor and sub-critic networks, and uses a Recurrent Neural Network with gate recurrent units (GRU) in cyclic layers to obtain the inputs of these networks. For the experiments we set $T=1$, i.e, the record of all items of the last round is viewed as the state.\footnote{We found out that training our algorithms for larger values of $T$ does not help to improve the performance.} We employ heuristic algorithms such as the Greedy Myopic Algorithm for exploration, i.e. we add these samples to the replay buffer before training.\\

\noindent \textbf{Experimental Parameters:} We use 1000 episodes for both training and testing, and there are 1000 steps in each episode. The valuation of the buyer in each round is drawn from the standard uniform distribution $U(0,1)$ and the costs of sellers follow a Gaussian distribution with mean $1/2$ and variance $1/2$. The size of the replay buffer is $10^5$, the discount factor $\gamma$ is $0.99$, and the rate of update of the target network is $10^{-3}$. The actor network and the critic network are trained via the \emph{Adam} algorithm, a gradient descent algorithm presented in \cite{kingma2014adam}, and the learning rates of these two networks are $10^{-4}$.  
Following the same idea as in \cite{lillicrap2015continuous}, we add Gaussian noise to the action outputted by the actor network, with the mean of the noise decaying with the number of episodes in the exploration.

\subsection*{Convergence of DDPG and IA(GRU)}

\begin{figure}
	\centering
	\subfigure[Loss of DDPG in training]{
		\label{Fig.sub.1}
		\includegraphics[scale=0.2]{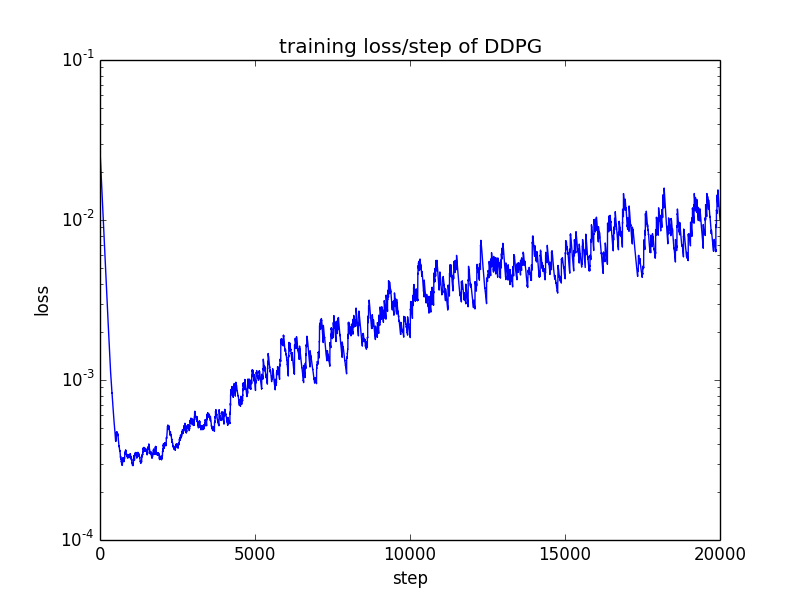}}
	\subfigure[Loss of IA(GRU) in training]{
		\label{Fig.sub.2}
		\includegraphics[scale=0.2]{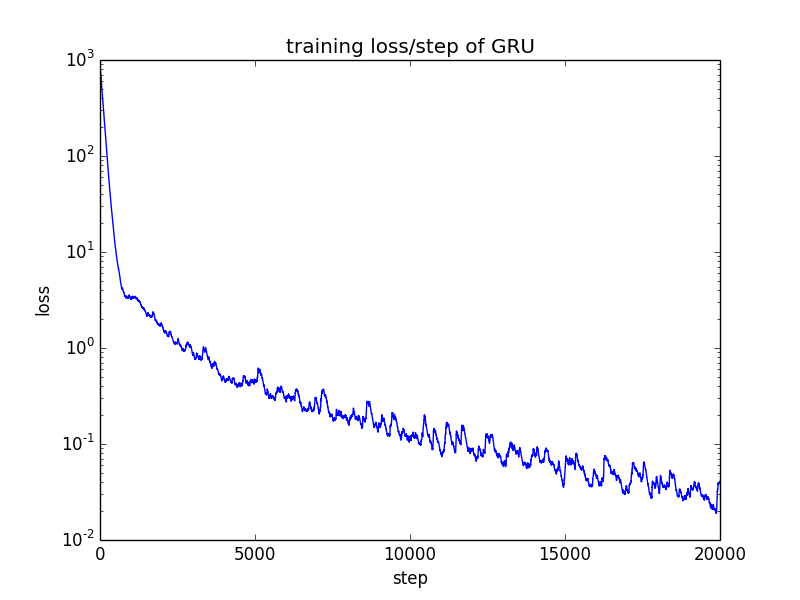}}
	\caption{Loss of DDPG and IA(GRU) in training for rational sellers.}
	\label{fig:3}
\end{figure}

First, to show the difference in the convergence properties of DDPG and IA(GRU), we train the algorithms for 200 sellers using the $\epsilon$-greedy strategy as the rationality model with variable costs for the sellers. Figure \ref{fig:2} shows the comparison between the rewards of the algorithms and Figure \ref{fig:3} shows the comparison in terms of the training loss with the number of steps. 

\begin{figure}\label{fig:2}
	\centering
	\subfigure[Rewards of DDPG in training]{
		\includegraphics[scale=0.2]{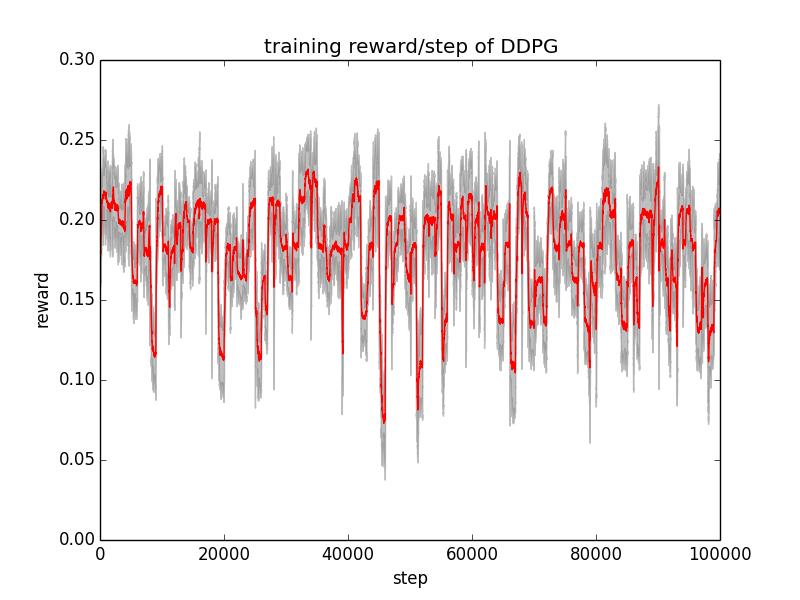}}
	\subfigure[Rewards of IA(GRU) in training]{
		\includegraphics[scale=0.2]{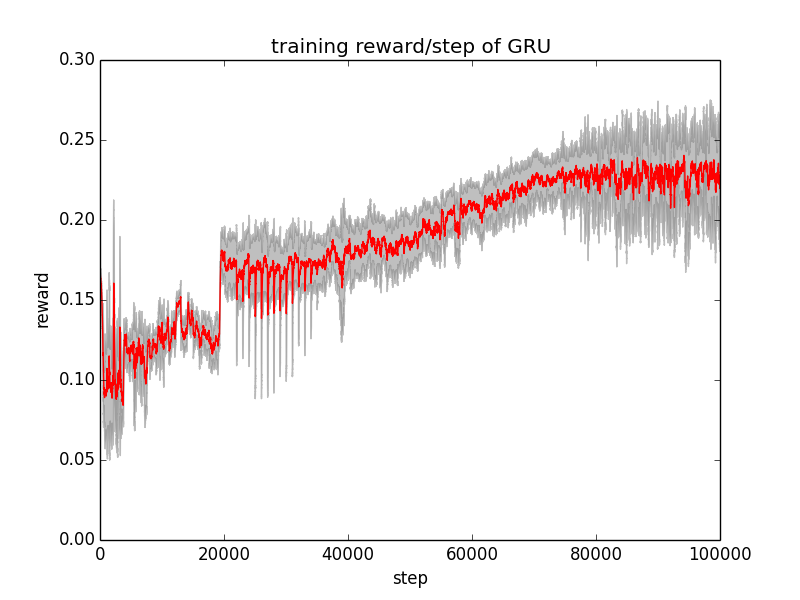}}
	\caption{Rewards of DDPG and IA(GRU) in training for rational sellers.}
	\label{fig:2}
\end{figure}

The gray band shows the variance of the vector of rewards near each step. From the figures, we see that DDPG does not converge, while IA(GRU) converges, as the training loss of the algorithm decreases with the number of steps. The convergence properties for the other rationality models are very similar.

\subsection*{Performance Comparison }

In this subsection, we present the revenue guarantees of IA(GRU) in the setting with $200$ sellers and how it fairs against the heuristics and DDPG, for either each rationality model separately, or for a heterogeneous pool of sellers, with a $1/4$-fraction of the sellers following each strategy. As explained in the previous page, we consider both the case of \emph{fixed} sellers and \emph{variable} sellers.\\

\begin{figure}
	\centering
	\includegraphics[scale=0.33]{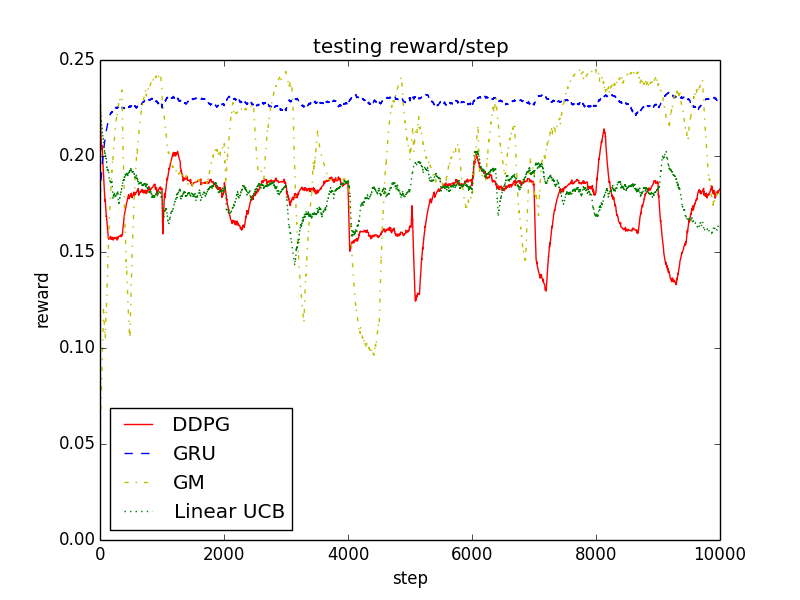}
	\caption{Rewards for fixed sellers and $\epsilon$-Greedy strategies.}
	\label{fig:4}
\end{figure}

\begin{figure}
	\centering
	\includegraphics[scale=0.33]{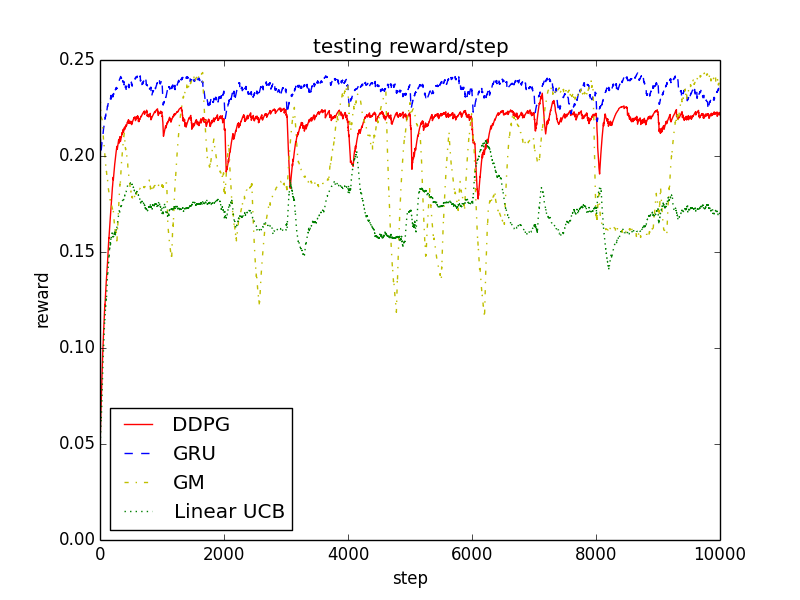}
	\caption{Rewards for fixed sellers and $\epsilon$-First strategies.}
	\label{fig:5}
\end{figure}

\begin{figure}
	\centering
	\includegraphics[scale=0.33]{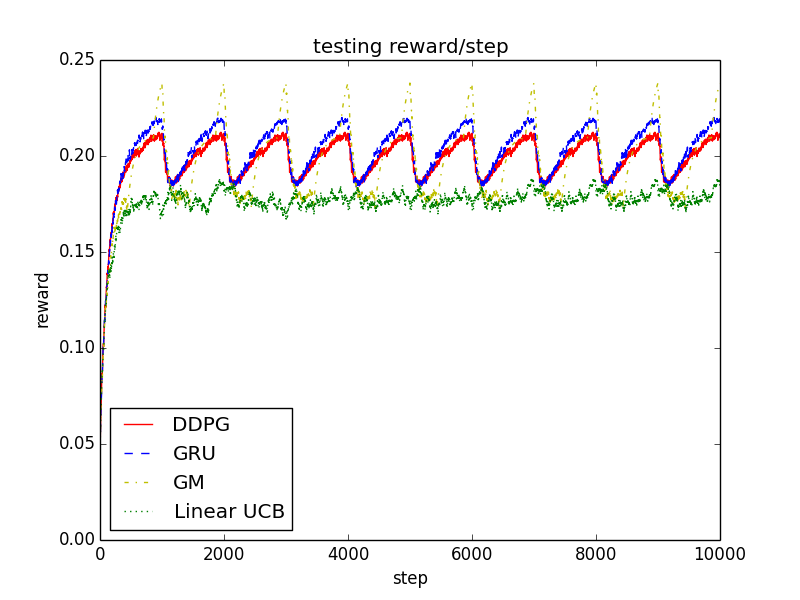}
	\caption{Rewards for fixed sellers and UCB1 strategies.}
	\label{fig:6}
\end{figure}

\begin{figure}
	\centering
	\includegraphics[scale=0.33]{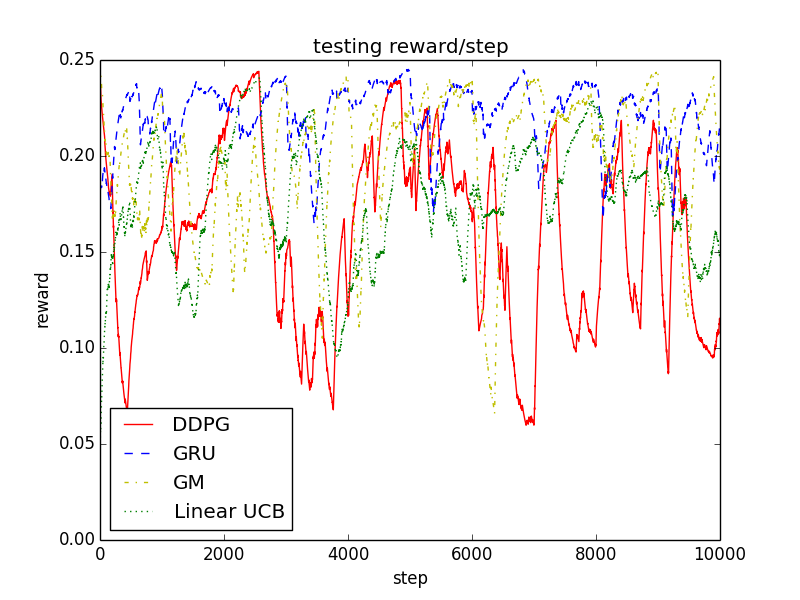}
	\caption{Rewards for fixed sellers and Exp3 strategies.}
	\label{fig:7}
\end{figure}

\begin{figure}
	\centering
	\includegraphics[scale=0.33]{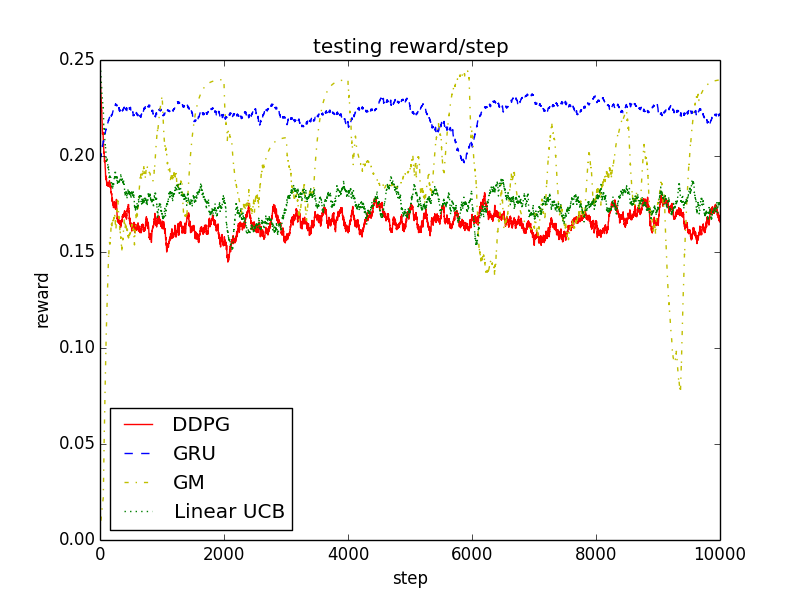}
	\caption{Rewards for fixed sellers and heterogeneous strategies.}
	\label{fig:8}
\end{figure}

\noindent \textbf{Performance Comparison for Fixed Sellers:} We show the performance of DDPG, IA(GRU), Greedy Myopic and Linear UCB on sellers using 
\begin{itemize}
	\item the $\epsilon$-Greedy strategy (Figure \ref{fig:4}),
	\item the $\epsilon$-First strategy (Figure \ref{fig:5}),
	\item the UCB1 strategy (Figure \ref{fig:6}),
	\item the Exp3 strategy (Figure \ref{fig:7}).
\end{itemize}
We also show the performance of the four different algorithms in the case of a heterogeneous population of sellers in Figure \ref{fig:8}.

Every point of the figures shows the reward at the corresponding step. We can conclude that the IA(GRU) algorithm is clearly better than the other algorithms in terms of the average reward on all rationality models. We also note that DPPG does not converge with 200 sellers and this is the reason for its poor performance.\\

\begin{figure}
	\centering
	\includegraphics[scale=0.33]{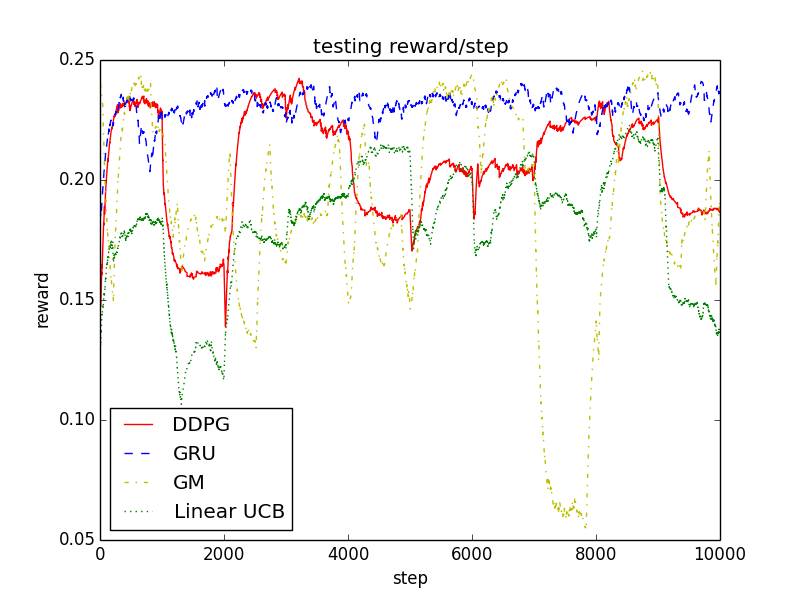}
	\caption{Rewards for variable sellers and $\epsilon$-Greedy strategies.}
	\label{fig:9}
\end{figure}

\begin{figure}
	\centering
	\includegraphics[scale=0.33]{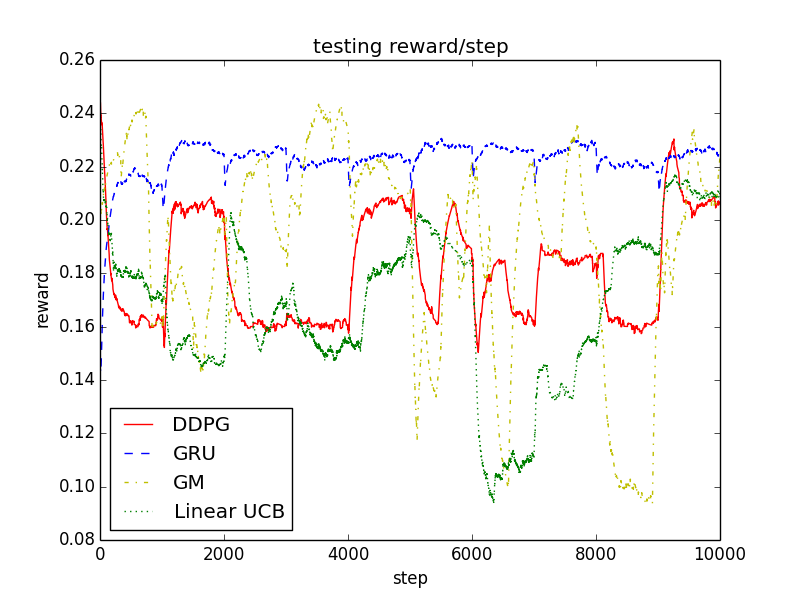}
	\caption{Rewards for variable sellers and $\epsilon$-First strategies.}
	\label{fig:10}
\end{figure}

\begin{figure}
	\centering
	\includegraphics[scale=0.33]{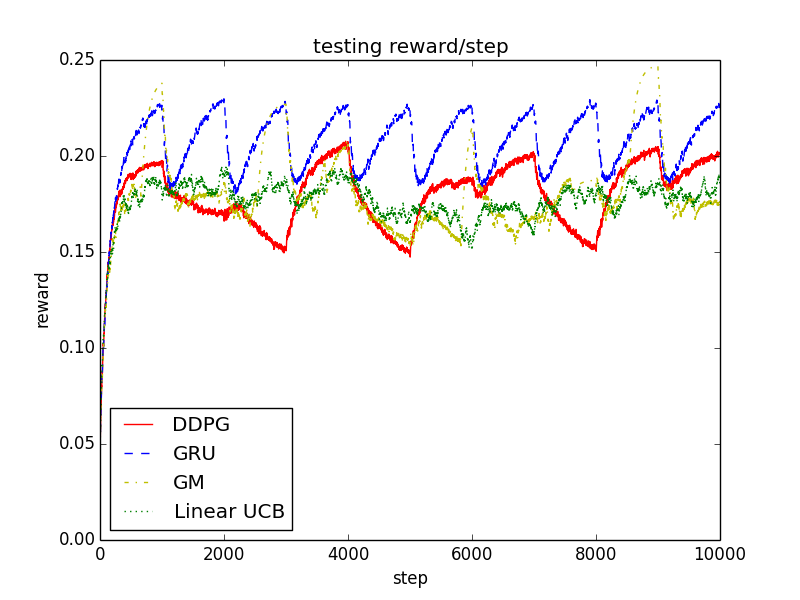}
	\caption{Rewards for variable sellers and UCB1 strategies.}
	\label{fig:11}
\end{figure}

\begin{figure}
	\centering
	\includegraphics[scale=0.33]{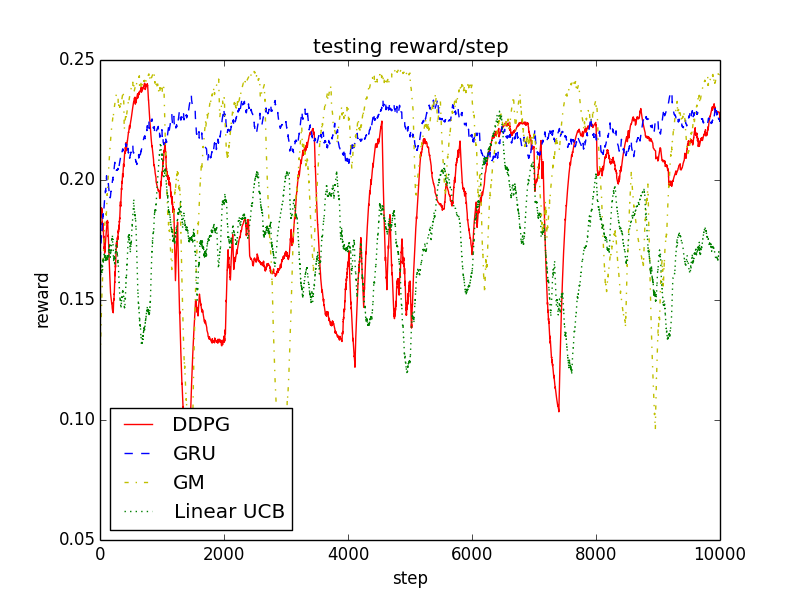}
	\caption{Rewards for variable sellers and Exp3 strategies.}
	\label{fig:12}
\end{figure}

\begin{figure}
	\centering
	\includegraphics[scale=0.33]{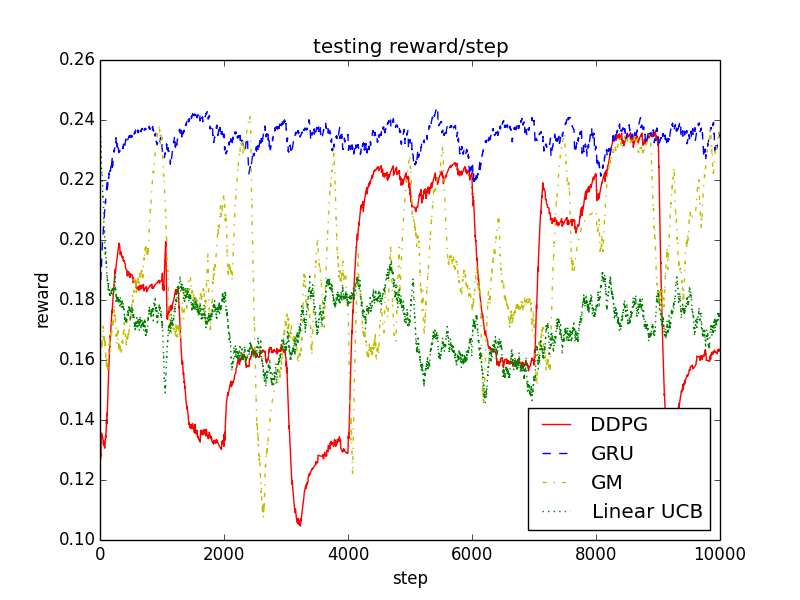}
	\caption{Rewards for variable sellers and heterogeneous strategies.}
	\label{fig:13}
\end{figure}

\noindent\textbf{Performance Comparison for Variable Sellers:} We show the performance of DDPG, IA(GRU), Greedy Myopic and Linear UCB on sellers using 
\begin{itemize}
	\item the $\epsilon$-Greedy strategy (Figure \ref{fig:9}),
	\item the $\epsilon$-First strategy (Figure \ref{fig:10}),
	\item the UCB1 strategy (Figure \ref{fig:11}),
	\item the Exp3 strategy (Figure \ref{fig:12}).
\end{itemize}
We also show the performance of the four different algorithms in the case of a heterogeneous population of sellers in Figure \ref{fig:13}.
Again here, we can conclude that the IA(GRU) algorithm clearly outperforms all the other algorithms in terms of the average reward on all rationality models. Also, IA(GRU) fairs better in terms of stability, as the other algorithms perform worse in the setting with variable sellers, compared to the setting with fixed sellers.

\subsection*{Scalability}
\begin{figure}
	\centering
	\includegraphics[scale=0.33]{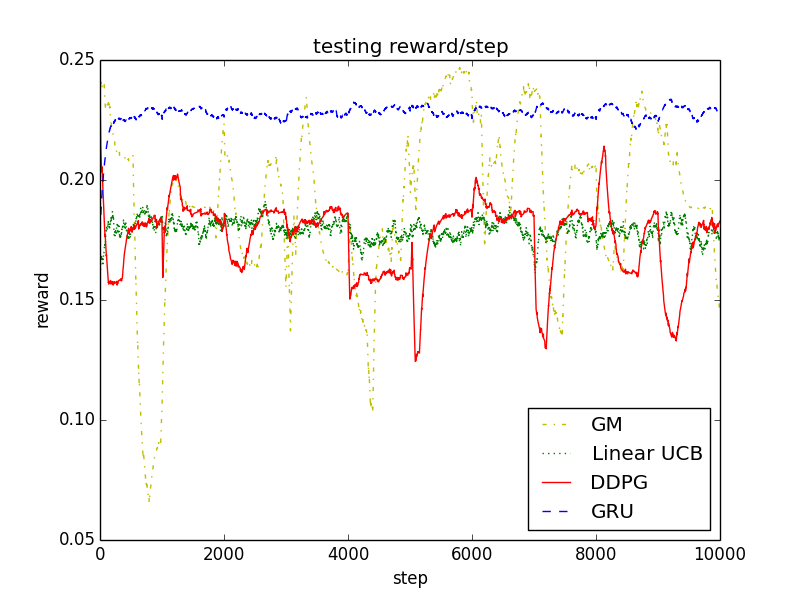}
	\caption{Rewards for 10.000 fixed sellers and $\epsilon$-Greedy strategies.}
	\label{fig:14}
\end{figure}

In this subsection, we present the revenue guarantees of IA(GRU) in the setting with $10000$ \emph{fixed} sellers and how it fairs against the heuristics and DDPG to show the scalability properties of IA(GRU) with the number of sellers. For IA(GRU) and DDPG, we will employ a simple ``scale-and-solve'' variant, since applying either of them directly to the pool of 10.000 sellers is prohibitive in terms of their running time. We design $50$ allocation sub-problems, consisting of $200$ sellers each, and divide the total number of impressions in $50$ sets of equal size, reserved for each sub-problem. We run IA(GRU) and DDPG algorithms in parallel for each sub-problem, which is feasible in reasonable time. For the heuristics, we run the algorithms directly on the large population of 10.000 sellers. The results for the case of $\epsilon$-Greedy seller strategies are show in Figure \ref{fig:14} (the results for other strategies are similar). We can see that even though we are applying a heuristic version, the performance of IA(GRU) is still clearly superior to all the other algorithms, which attests to the algorithm being employable in larger-case problems as well.

\section{Conclusion}

In this paper, we employed a reinforcement mechanism design framework for solving the impression allocation problem of large e-commerce websites, while taking the rationality of sellers into account. Inspired by recent advances in reinforcement learning, we designed a deep reinforcement learning algorithm which outperforms several natural heuristics under different realistic rationality assumptions for the sellers in terms of the generated revenue, as well as state-of-the-art reinforcement learning algorithms in terms of performance and convergence guarantees. 

Our algorithm can be applied to other dynamical settings for which the objectives are similar, i.e. there are multiple agents with evolving strategies, with the objective of maximizing a sum of payments or the generated revenue of each agent. It is an interesting future direction to identify several such concrete settings and apply our algorithm (or more generally our framework), to see if it provides improvements over the standard approaches, as it does here.

\begin{acks}
Qingpeng Cai and Pingzhong Tang were supported in part by the National Natural Science Foundation of China Grant 61561146398, a Tsinghua University Initiative Scientific Research Grant, a China Youth 1000-talent program and Alibaba Innovative Research program. Aris Filos-Ratsikas was supported by the ERC Advanced Grant 321171 (ALGAME).

\end{acks}

\clearpage
\bibliographystyle{ACM-Reference-Format}
\balance
\bibliography{reinforcement} 

\end{document}